\documentstyle[prb,aps,multicol,epsf]{revtex}

\def\Bildeins{\epsfbox{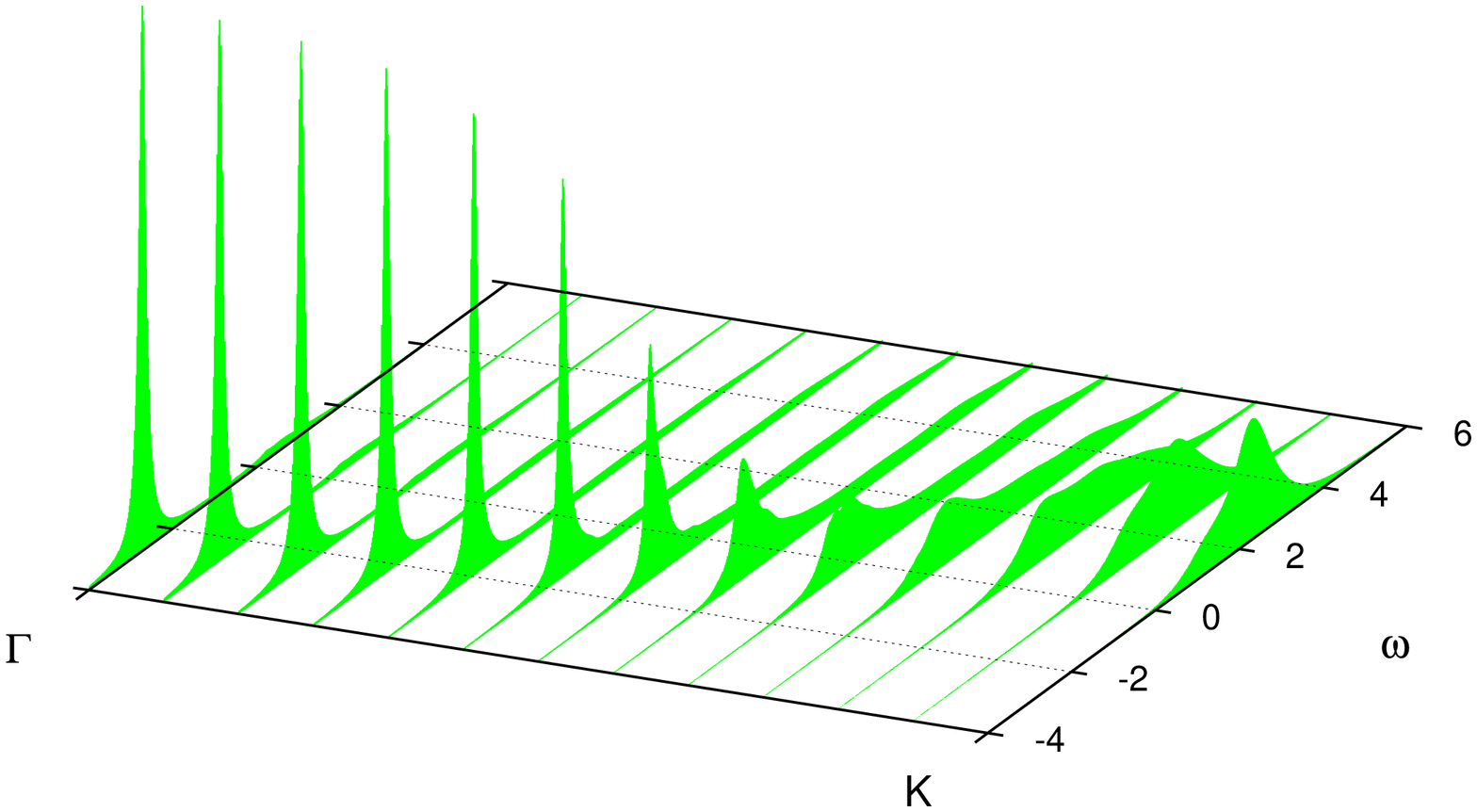}}
\def\Bildzwei{\epsfbox{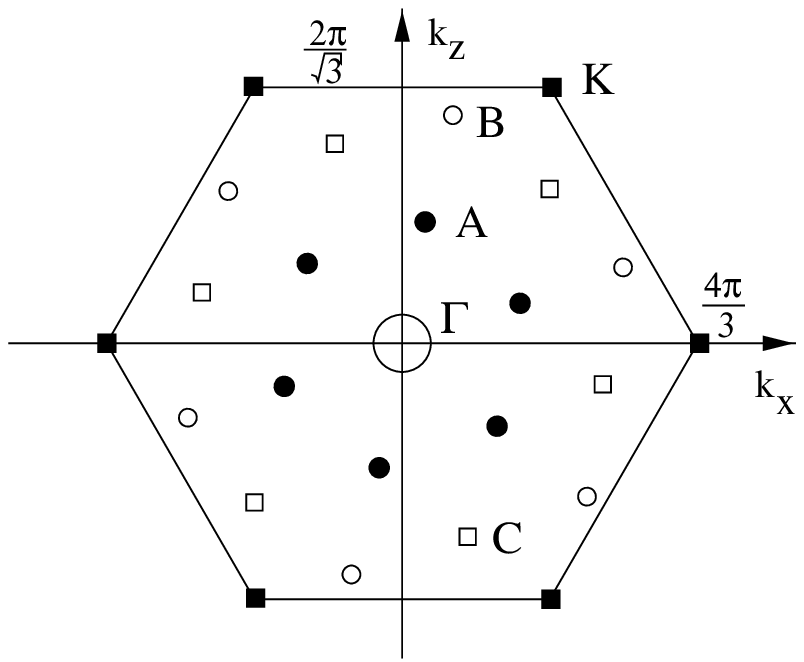}}
\def\Bilddreia{\epsfbox{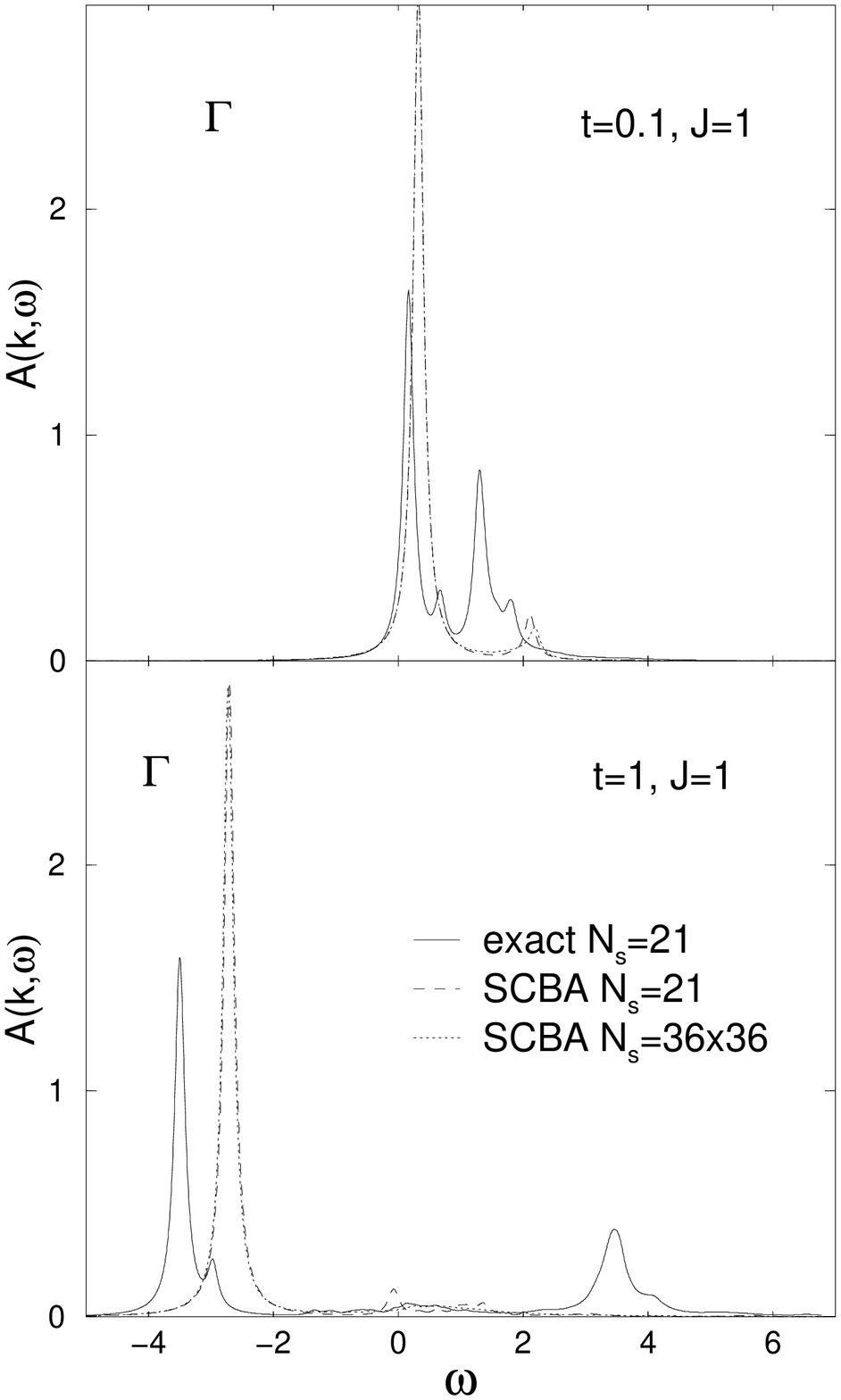}}
\def\Bilddreib{\epsfbox{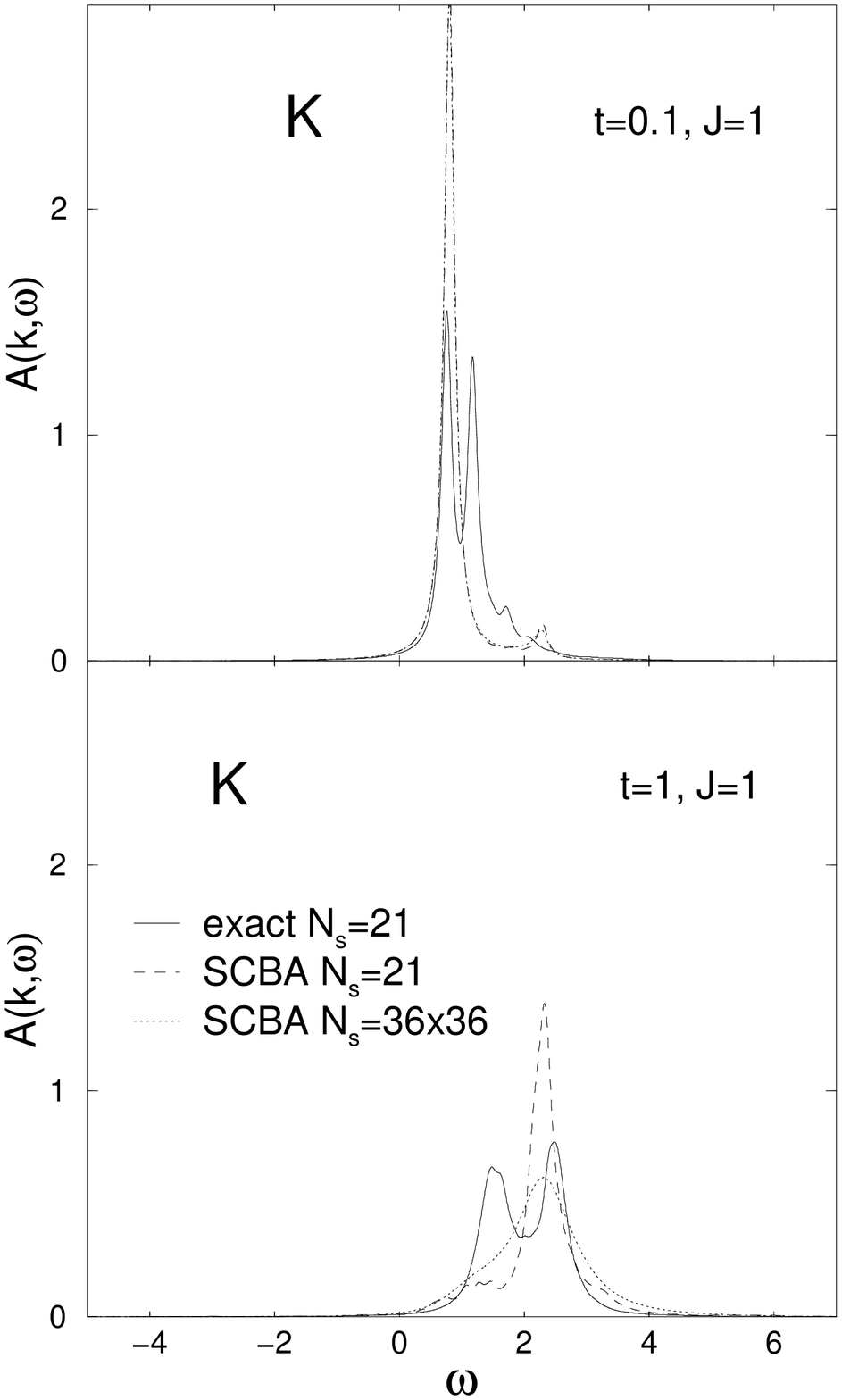}}
\def\Bildvier{\epsfbox{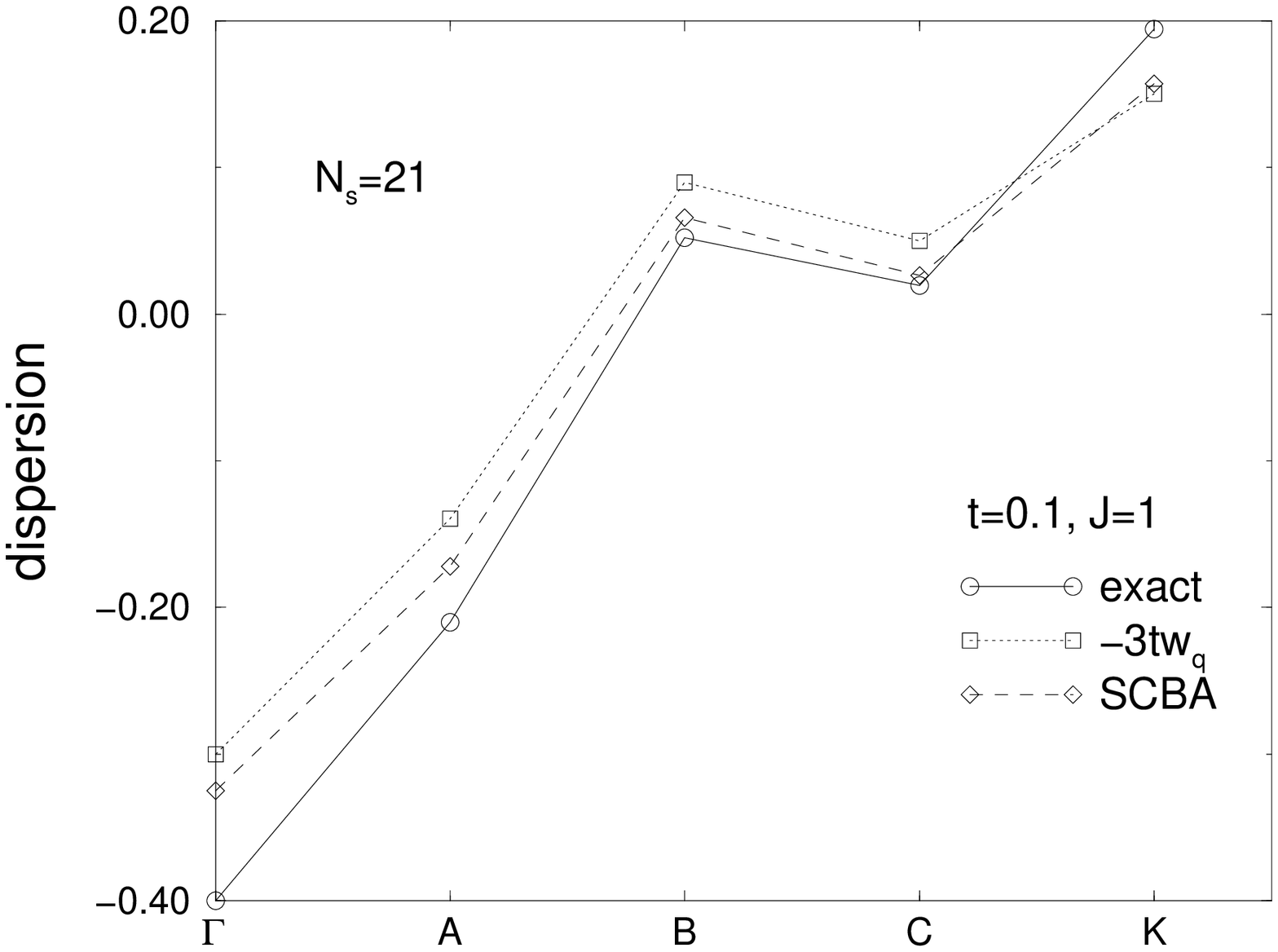}}
\def\Bildfunf{\epsfbox{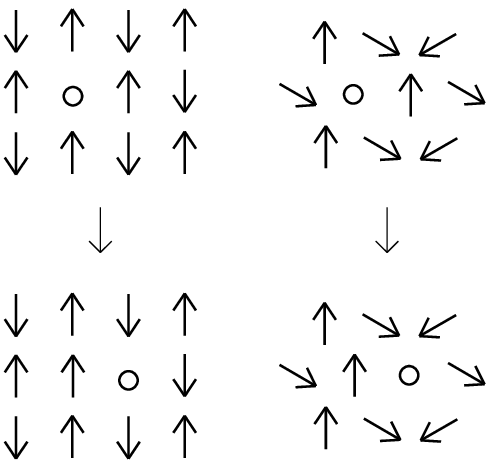}}

\begin{document}

%%%%%%%%%%%%%%%%%%%%%%%%%%%%%%%%%%%%%%%%%%%%%%%%%%%%%%%%%%%%%%%%%%%%%%%%%%%%%%%

\title{Effective Hamiltonians for holes in antiferromagnets:
a new approach to implement forbidden double occupancy}

\author{W. Apel$^{a}$, H.-U. Everts$^{b}$ and U. K\"orner$^{b}$ }

\address{$^{a}$ Physikalisch-Technische Bundesanstalt, Bundesallee 100,
D 38116 Braunschweig, Germany\\
$^{b}$ Institut f\"ur Theoretische Physik, Universit\"at Hannover,\\
Appelstr. 2, D 30167 Hannover, Germany \\
}

\date{\today}
\maketitle

\begin{abstract}
A coherent state representation for the electrons of ordered antiferromagnets
is used to derive effective Hamiltonians for the dynamics of holes in such
systems. By an appropriate choice of these states, the constraint of forbidden
double occupancy can be implemented rigorously. Using these coherent states,
one arrives at a path integral representation of the partition function
of the systems, from which the effective Hamiltonians can be read off.
We apply this method to the t-J model on the square lattice and on the
triangular lattice. In the former case, we reproduce the well-known
fermion-boson Hamiltonian for a hole in a collinear antiferromagnet. We
demonstrate that our method also works for non-collinear antiferromagnets
by calculating the spectrum of a hole in the triangular antiferromagnet in
the self-consistent Born approximation and by comparing it with numerically
exact results.
\end{abstract}


\begin{multicols}{2}
\narrowtext

\section{Introduction}
It is widely recognized that an accurate description of the dynamics of
holes in antiferromagnetically (AF) ordered materials is an important
first step towards an understanding of the essential physics of the cuprate
superconductors.
Consequently, numerous investigations
\cite{Dagotto94,Tohyama96,Xiang96,Hayn96,Sheng96,Vojta96,Brinkman70,Trugman90,%
Li89,Pecher95,Shankar90,Wiegmann88}
have dealt with the problem of a single hole in a square-lattice
antiferromagnet.
Most of these studies are based on an effective Hamiltonian ${\cal H}_{\Box}$
\cite{Kane89,Marsiglio91} which describes the hole as a spinless fermion
which is coupled to bosons representing the collective excitations (magnons)
of the antiferromagnetic background.
The structure of this effective Hamiltonian reflects the two-sublattice
structure of the square lattice AF.
When the hole hops between nearest neighbour lattice sites, it necessarily
disturbs the magnetic order.
Thus, any move of the hole requires absorption or emission of a magnon.
Derivations of ${\cal H}_{\Box}$ from the t-J Hamiltonian have revealed
the approximate nature of this effective Hamiltonian \cite{Barentzen94},
and to obtain the spectrum of a hole from ${\cal H}_{\Box}$
further approximations are necessary.
Most often, the fermion-magnon coupling is treated in the self-consistent
Born approximation (SCBA) which was already used in the original studies
by Kane et al. \cite{Kane89} and Marsiglio et al. \cite{Marsiglio91}.
Surprisingly, the approximate spectra obtained in this manner agree excellently
with the spectra of small clusters obtained by numerically exact
diagonalization of the $t-J$ Hamiltonian for these clusters.

Apart from its relevance to the physics of cuprate superconductors, the
dynamics of holes in ordered antiferromagnets is a highly nontrivial
problem of the physics of strongly correlated electron systems, and it is
therefore of fundamental theoretical interest. Here, we shall present
a derivation of effective Hamiltonians for holes in AFs which is not
restricted to any particular type of magnetic order. Our method does not
provide us with a unique effective Hamiltonian for an arbitrary number of
holes. Rather the cases of a single hole, two holes and in general an
arbitrary but fixed number of holes have to be treated separately. While
our presentation remains general, we shall confine ourselves to the
single-hole problem in order to demonstrate our method's practical
applicability. In
particular, we shall derive and evaluate the effective Hamiltonian
${\cal H}_{\triangle}$ for a single hole in the triangular AF.

\section{Effective Lagrangian}
For definiteness, we consider the $t-J$ Hamiltonian,
\begin{eqnarray}
{\cal H}&=& -t \sum_{\stackrel {<{\bf r},{\bf r'}>}{\sigma}}
\hat{P} (c_{{\bf r},\sigma}^{\dagger} c_{{\bf r'},\sigma}^{} + {\it h.c.})
\hat{P} \nonumber\\
&&+J\sum_{<{\bf r},{\bf r'}>} \hat{\bf S}_{\bf r}\cdot \hat{\bf S}_{\bf r'}\, .
\label{t-J}
\end{eqnarray}
$t$ and $J$ are the single particle hopping matrix element and the exchange
constant, respectively, the operator $\hat{P}$ projects onto states in which
each of the $N_s$ lattice sites is either empty or singly occupied, and
$\hat{\bf S}_{\bf r}=\frac{1}{2} c_{{\bf r},\alpha}^{\dagger}
\mbox{\boldmath $\sigma$}_{\alpha,\beta} c_{{\bf r},\beta}^{}$
is the electron spin expressed in terms of creation and annihilation operators,
$c^{\dagger}_{{\bf r},\alpha}$ and $c^{}_{{\bf r},\alpha}$,
for fermion states at site ${\bf r}$ with spin projection $\alpha$.
The sums in (\ref{t-J}) run over pairs $<\, , \,>$ of nearest neighbour sites
of the lattice.

In order to cast the partition function ${\cal Z}={\it tr}\exp\{-\beta
{\cal H}\}$ into the form of a path integral, we introduce the following coherent states \cite{Wiegmann88}
for each lattice site ${\bf r}$ (we omit the site index wherever this does not
lead to ambiguities):
\begin{eqnarray}
|\omega\!>&=&e^{\left[\eta c-c^{\dagger}\eta^*\right]} |\Omega\!>,\nonumber\\
|\Omega\!>&=&c^{\dagger} |0\!>, \nonumber\\
 c^{\dagger}&=&
 e^{-i\phi/2}\cos\frac{\theta}{2} e^{-i\psi/2}c_{\uparrow}^{\dagger}
+e^{ i\phi/2}\sin\frac{\theta}{2} e^{-i\psi/2}c_{\downarrow}^{\dagger}.
\label{coh}
\end{eqnarray}
Here, $\eta$, $\eta^*$ are Grassmann variables, and $|0\!>$ is the vacuum
with no spins present.
Obviously, $|\Omega\!>$ is a one-fermion state with spin in the direction
$(\theta,\phi)$ on the sphere.
The inclusion of a phase factor which depends on the third Euler angle $\psi$
is necessary, if $|\Omega\!>$ and $\eta$ are to satisfy the following
requirements: ({\it i}) $|\Omega\!>$ is to transform covariantly under
{\it all} $SU(2)$ operations, while ({\it ii}) $\eta$ remains invariant
under these operations.
As will be seen shortly, both requirements appear natural in view of
the physical interpretation of the variables $\Omega=(\psi,\theta,\phi)$
and $\eta$.
It is easily checked that the states $|\omega\!>$ {\it include}
the empty and the spin-up and spin-down states of the fermion, but {\it
exclude} the doubly occupied state. In fact, the states $|\omega\!>$ are
complete in this three-dimensional space,
\begin{eqnarray}
&&\int d\omega\, |\omega\!><\!\omega|:=\nonumber\\
&& \int d\Omega\,\frac{1}{2}\int d\eta^{*} d\eta\,
e^{-\eta^{*}\eta}|\omega\!><\!\omega|=\nonumber\\
&&\int d\Omega\,\frac{1}{2}\int d\eta^{*} d\eta\,e^{-2\eta^{*}\eta}
\left\{\eta \,\eta^* |0\!><\!0|+|\Omega\!><\!\Omega|\,\right\}\nonumber\\
&&={\bf 1}_3
\label{comp}
\end{eqnarray}
where $\int d\Omega \cdots = \frac{1}{8\pi^2} \int_{0}^{4\pi}d\psi
\int_{0}^{\pi}d\theta \sin{\theta} \int_{0}^{2\pi}d\phi \cdots$.
The Euler angles $\Omega$ describe the spin degrees of freedom and the
Grassmann variables $\eta$ and $\eta^{*}$ a spinless fermionic hole.
Thus the requirement ({\it ii}) is cogent.
With these coherent states, the standard steps towards a path integral
representation of the partition function yield
\begin{eqnarray}
{\cal Z}_M&=&\int {\cal D}_M[\Omega]\int {\cal D}_M[\eta^*,\eta^{}]\nonumber\\
&&\prod_{\stackrel{\tau=1}{{\bf r}}}^M
\left(<\!\Omega_{\tau-1,{\bf r}}|\Omega_{\tau,{\bf r}}\!>\frac{1}{2}
e^{-2\eta_{\tau,{\bf r}}^{*}\eta_{\tau,{\bf r}}^{}}\right)\nonumber\\
&& \exp\left\{\sum_{\tau,{\bf r}} \frac{\eta_{\tau-1,{\bf r}}^*\eta_{\tau,{\bf
r}}^{}}{<\!\Omega_{\tau-1,{\bf r}}|\Omega_{\tau,{\bf r}}\!>} +
\right.\nonumber\\
&& \left. \sum_{\tau}
\ln\left[1+\frac{<\!\omega_{\tau-1}| e^{-\Delta\tau {\cal H}} -
1|\omega_{\tau}\!>}{<\!\omega_{\tau-1}|\omega_{\tau}\!>} \right]\right\},
\label{Z}
\end{eqnarray}
with ${\cal D}_M[\Omega]=\prod_{\tau=1,{\bf r}}^M d\Omega_{\tau,{\bf r}}$ and
${\cal D}_M[\eta^*,\eta]=\prod_{\tau=1,{\bf r}}^M d\eta^{*}_{\tau,{\bf r}}d\eta^{}_{\tau,{\bf r}}$.
Here, sums and products run over the $M$ (imaginary) time slices of the Trotter
decomposition of $\exp(-\beta {\cal H})$ (subscript $\tau$) and over the $N_s$
lattice sites (subscript ${\bf r}$).
$\Delta\tau=\beta/M$ is the width of the time slices.
The boundary conditions are $\Omega_{0,{\bf r}}=\Omega_{M,{\bf r}}$ and
$\eta^{(*)}_{0,{\bf r}}=-\eta^{(*)}_{M,{\bf r}}$.
The final step in the conversion of ${\cal Z}_M$ into a path integral would be
to take the continuum limit $M \rightarrow \infty$ in the time direction.
This meets with difficulties.
After the expansion of the second exponential in (\ref{Z}) to linear order in
$\Delta\tau$, a factor
$\prod_{\tau,{\bf r}}e^{-\eta_{\tau,{\bf r}}^*\eta_{\tau,{\bf r}}^{}}$
remains in the integrand which is therefore {\it ill} defined in the limit
$M\rightarrow \infty$. However, as we shall show now, the individual
summands ${\cal Z}^{n-hole}$ in the decomposition
\begin{equation}
{\cal Z}_M=
{\cal Z}_{M}^{0-hole}+{\cal Z}_{M}^{1-hole}+\ldots+{\cal Z}_{M}^{n-hole}+\ldots
\label{dec}
\end{equation}
of the partition function ${\cal Z}_M$ into
$0$-, $1$-, $\ldots n$-hole contributions are {\it well} defined in the
$\tau$-continuum limit. Since in this work we are interested in the
one-hole effective Hamiltonian, which can be extracted from ${\cal Z}^{1-hole}$,
we confine our attention to the first two terms of (\ref{dec}).
The crucial elements in this decomposition are the weight factors
$e^{-2\eta_{\tau,{\bf r}}^{*}\eta^{}_{\tau,{\bf r}}}
=1-2\eta^{*}_{\tau,{\bf r}} \eta^{}_{\tau,{\bf r}}$
in the integrand of (\ref{Z}). From (\ref{comp}) it follows that
the identity in such a factor projects onto configurations with a hole at
$(\tau,{\bf r})$ while the second term, $-2\,\eta^{*}_{\tau,{\bf r}} \eta^{}_{\tau,{\bf r}}$, projects onto
configurations with an electron at $(\tau,{\bf r})$. Thus, by retaining in
the integrand of (\ref{Z}) only the term
$\prod_{\tau,{\bf r}}\left(-2\eta^{*}_{\tau,{\bf r}}\eta^{}_{\tau,{\bf r}}\right)$
of the expansion of
$\prod_{\tau,{\bf r}}\left(1-2\eta^{*}_{\tau,{\bf r}}\eta^{}_{\tau,{\bf r}}\right)$,
one obtains ${\cal Z}^{0-hole}$, the partition function with no hole at all
which is, of course, the partition function of the Heisenberg Hamiltonian.
Similarly, by retaining those terms of the expansion which contain for each
time slice $\tau$ just $N_s-1$ factors
$-2\eta^*_{\tau,{\bf r}}\eta^{}_{\tau,{\bf r}}$ one obtains
${\cal Z}^{1-hole}_M$.
(The number of holes remains constant in time.)
Using the identity
\begin{equation}
\prod_{\tau,{\bf r}} \left. \left( \frac{1}{2} e^{-2\eta_{\tau,{\bf r}}^{*}
\eta^{}_{\tau,{\bf r}}} \right) \right|_{1-hole} =
\frac{1}{2^M} \prod_{\tau,{\bf r}} \left. \left( e^{- \eta^{*}_{\tau,{\bf
r}} \eta^{}_{\tau,{\bf r}}} \right) \right|_{1-hole},
\end{equation}
which holds between the one-hole projectors on both sides,
we can now cast ${\cal Z}^{1-hole}_M$ into the form
\begin{eqnarray}
{\cal Z}^{1-hole}_M&=&
\frac{1}{2^M} \int{\cal D}_M[\Omega] \int{\cal D}_M[\eta^*,\eta]\nonumber\\
&&\prod_{\stackrel{\tau=1}{{\bf r}}}^M
\left(<\!\Omega_{\tau-1,{\bf r}}|\Omega_{\tau,{\bf r}}\!> \right)\nonumber\\
&&\exp\left\{-\sum_{\tau,{\bf r}} \left(\eta_{\tau,{\bf r}}^*\eta_{\tau,{\bf r}}^{}-
\frac{\eta_{\tau-1,{\bf r}}^*\eta_{\tau,{\bf r}}^{}}{<\!\Omega_{\tau-1,{\bf r}}|
\Omega_{\tau,{\bf r}}\!>}\right)\right. + \nonumber\\
&& \,\, \left. \left.\sum_{\tau}
\ln\left[1+\frac{<\!\omega_{\tau-1}| e^{-\Delta\tau {\cal H}} -
1|\omega_{\tau}\!>}{<\!\omega_{\tau-1}|\omega_{\tau}\!>}
\right]\right\}\right|_{1-hole}\!\!\!\!\!.\nonumber\\
\label{ZH1M}
\end{eqnarray}
The additional factor of $\frac{1}{2}$ associated with each time slice
$\tau$ finds a natural explanation:
It compensates for the integrations over the redundant spin variable at the
position of the hole \cite{MacDonald97}.
Obviously, the integrand of (\ref{ZH1M}) is well defined in the
$\tau$-continuum limit $M \rightarrow \infty$. Performing this limit, one obtains
\begin{equation}
{\cal Z}^{1-hole}\,\,
=\,\,\left.\int {\cal D}[\Omega]\int {\cal D}[\eta^{*},\eta]\,\,
e^{\,\int^{\beta}_0 d\tau {\cal L}}\right|_{1-hole}
\label{ZH1}
\end{equation}
with the classical Lagrangian for one hole
\begin{eqnarray}
{\cal L}&=&\sum_{\bf r}\left[1-
\eta_{\bf r}^{*}(\tau)\eta_{\bf r}^{}(\tau)\right]<\!\Omega_{\bf r}(\tau)|
\dot{\Omega}_{\bf r}(\tau)\!> \nonumber\\
&&+\sum_{\bf r} \eta^{*}_{\bf r}(\tau) \dot{\eta}_{\bf r}^{}(\tau)
\,-\,<\omega(\tau)|{\cal H}|\omega(\tau)>_{1-hole}.
\label{LH1}
\end{eqnarray}
Here, $|\omega(\tau)\!>=\prod_{\bf r}|\omega_{\bf r}(\tau)\!>$, and
$<\omega(\tau)|{\cal H}|\omega(\tau)>_{1-hole}$ denotes that part of the
expectation value which is bilinear in the Grassmann fields $\eta^{*},\eta$.

It should be clear from these considerations that in order to determine the
effective interaction between a pair of holes, one will have to analyze the
2-hole contribution to ${\cal Z}_M$. In calculating it from the 1-hole
contribution by integrating over the spin degrees of freedom, one would miss a
part of this interaction.

Evaluating (\ref{LH1}) with the states (\ref{coh}) and the Hamiltonian
(\ref{t-J}) of the $t-J$ model on an arbitrary lattice,
we obtain ${\cal L}= {\cal L}_{kin}+{\cal L}_t+{\cal L}_J$, where
(we omit total time derivatives)
\begin{eqnarray}
&&{\cal L}_{kin}=\sum_{\bf r} \left[\eta^*_{\bf r} \dot{\eta}^{}_{\bf r}
+\frac{i}{2}\eta^*_{\bf r}\eta^{}_{\bf r}\dot{\psi}^{}_{\bf r}\right.\nonumber\\
&&\qquad\qquad-\left.\frac{i}{2}\left(1-\eta^{*}_{\bf r}\eta^{}_{\bf r}\right)
\dot{\phi}_{\bf r}\cos(\theta_{\bf r})\right] \; ,
\label{Lkin}
\end{eqnarray}
\begin{eqnarray}
&&{\cal L}_t=-t \,\sum_{<{\bf r,r'}>}\left\{\eta^*_{\bf r'}\eta^{}_{\bf r}
     e^{-\frac{i}{2}(\psi_{\bf r'}-\psi_{\bf r}^{})}\right.\nonumber\\
&&\left[ \cos\left(\frac{\phi_{\bf r}-\phi_{\bf r'}}{2}\right)
       \cos\left(\frac{\theta_{\bf r}-\theta_{\bf r'}}{2}\right)\right.
\nonumber\\
&&\left.\left.+\,\,i\sin\left(\frac{\phi_{\bf r}-\phi_{\bf r'}}{2}\right)
       \cos\left(\frac{\theta_{\bf r}+\theta_{\bf r'}}{2}\right)\right]\,
        +\,({\bf r}\leftrightarrow {\bf r'}) \right\} 
\end{eqnarray}
and
\begin{equation}
{\cal L}_J=-J\sum_{<{\bf r,r'}>}(1-\eta^{*}_{\bf r}\eta^{}_{\bf r}\,\, -\,\, \eta^{*}_{\bf r'}\eta^{}_{\bf r'})\;
{\bf S}_{\bf r}\cdot{\bf S}_{\bf r'} \; .
\label{Lj}
\end{equation}
The Heisenberg term ${\cal L}_J$ accounts for the interaction between spins
${\bf S}=\frac{1}{2} (\sin\theta \cos\phi, \sin\theta \sin\phi,\cos\theta )$
at sites where no hole is present.
${\cal L}_t$ describes the hopping of the hole; the hopping energy depends
on the state of the spins.
Finally, ${\cal L}_{kin}$ contains the kinetic terms of a spinless hole
($\eta^{*}\dot{\eta}$), and a spin $\frac{1}{2}$, i.e. $\dot{\phi}\cos(\theta)$.
In addition, two terms couple the hole to the angular degrees of freedom.
The second cancels the kinetic term of the spin at sites where a hole is
present.
The first has the form of a gauge interaction with the field $\psi$.
It is tempting to use a parameterization in which the factor
$\exp(\frac{i}{2}\psi)$ is absorbed in $\eta$, since then, all the $\psi$
dependent terms in ${\cal L}$ disappear.
However, after such a gauge transformation, $\eta$ is no longer invariant
under $SU(2)$ operations, as explained above in connection with the choice
of the coherent states (\ref{coh}).
Since this has unwanted consequences, we do not follow this route.

In summary, our procedure for obtaining effective Hamiltonians consists of
the following steps:
first we represent the partition function of the initial quantum Hamiltonian
as a discrete-time path integral; then, we identify that part of the partition
function that corresponds to the number of holes we wish to consider; in this
part, we then perform the continuum limit to obtain the classical Lagrangian;
finally, we translate back to an effective quantum Hamiltonian.

The Lagrangian ${\cal L}$, (\ref{Lkin}-\ref{Lj}), still represents the full
non-linear problem of a hole interacting with the spin background. In the
following, we shall confine ourselves to a spin wave expansion around the 
classical spin groundstate of ${\cal L}$.
In this picture, the hole moves in a spin background described by angular fields
which deviate little from their groundstate. The spin wave expansions have, of
course, to be performed separately for the case of collinear spin order (square
lattice AF) and for the case of planar spin order (triangular lattice AF).
We shall briefly rederive the quantum Hamiltonian, ${\cal H}_{\Box}$, which is
well known for the case of collinear spin order. For the case of planar spin
order, we obtain a significantly different effective Hamiltonian
${\cal H}_{\triangle}$. To
check the validity of the approximations that lead to ${\cal H}_{\triangle}$
we shall calculate the one hole spectral properties that follow from
${\cal H}_{\triangle}$ and compare them with results obtained by numerically 
exact diagonalization of the $t-J$ model for small finite lattice cells.

\section{One hole on a square lattice}
The two sublattice classical AF order is reproduced by assigning the value
$\pi/2$ to $\theta_{\bf r}$ and the values $\phi_{\bf r}^{(0)}=\pm\pi/2$ to
$\phi_{\bf r}$ for ${\bf r}$ from the A or B lattice, respectively.
The deviations from the ordered groundstate, $x_{\bf r}$ and $p_{\bf r}$
defined by
\begin{equation}
 \phi^{}_{\bf r}=\phi^{(0)}_{\bf r}+ \sqrt{2}x_{\bf r} \mbox{\hspace{1cm}}
\theta_{\bf r}=\frac{\pi}{2}+\sqrt{2} p_{\bf r} \;,
\label{coorsq}
\end{equation}
are canonically conjugate harmonic oscillator fields as can be seen from
${\cal L}_{kin}$.
Now, we expand ${\cal L}$ up to quadratic order in the amplitudes 
$a=(x-ip)/\sqrt{2}$ and $a^*$
and keep in the quadratic term only the leading zero--hole contribution.
Then, the spin and the hole degrees of freedom decouple in ${\cal L}_{kin}$.
We redefine the hole field by
$h\,\equiv\,\eta\,e^{\frac{i}{2}(\psi - \phi^{(0)})}$.
Then, all the phases in the hopping term in the path integral become equal;
the field $\psi$ disappears from ${\cal L}$ and is integrated out.
Finally, we translate back to operator form and get
\begin{eqnarray}
 {\cal H}_{\Box}&=&-t\sum_{<\!{\bf r}, {\bf r'}\!>}
 \left[h^{\dagger}_{\bf r'}h^{}_{\bf r} \, i \,
 \left(a^{\dagger}_{\bf r}-a^{}_{\bf r'}\right)+h.c.\right] \nonumber\\
&&+\frac{J}{2} \sum_{<\!{\bf r}, {\bf r'}\!>}
 \left(a^{\dagger}_{\bf r}a^{}_{\bf r} +a^{\dagger}_{\bf r'}a^{}_{\bf r'}
  -a^{\dagger}_{\bf r}a^{\dagger}_{\bf r'}-a^{}_{\bf r'}a^{}_{\bf r}\right)
\nonumber\\
&&+J \sum_{\bf r} \left(h^{\dagger}_{\bf r}h^{}_{\bf r}-\frac{1}{2}\right).
\end{eqnarray}
${\cal H}_{\Box}$ is, up to a canonical transformation
($a_{\bf r} \rightarrow - i a_{\bf r}$) identical with the Hamiltonian
considered in \cite{Marsiglio91}.
The last term in ${\cal H}_{\Box}$, containing the hole number,
accounts for the breaking of four bonds per hole in the classical groundstate.

\section{One hole in the triangular lattice}
In the classical limit, the groundstate of the $t-J$ model at half filling is
the well known planar ${120}^{\small o}$ spin structure:
$\theta^{(0)}_{\bf r}=\pi/2$ for all
${\bf r}$ and $\phi^{(0)}_{\bf r}=\frac{2\pi}{3},0,-\frac{2\pi}{3}$ for
the A, B and C sublattice, respectively.
As in the case of the square lattice, we describe the deviations from this
ordered groundstate by the spin wave amplitudes $x_{\bf r}$ and $p_{\bf r}$
which are canonically conjugate harmonic oscillator fields.

Proceeding as in the square lattice case, i.e. expanding to second order
in the amplitudes $a$ and $a^*$ and integrating over $\psi$, one would be left
with terms in ${\cal L}_t$ that couple the hole motion to zero energy spin modes.
But these modes are uniform spin rotations which are exact symmetries
of the initial model (\ref{t-J}). Therefore, they cannot appear in any order
of a perturbation expansion.
In the Appendix, we trace this apparent inconsistency back to the integration
over the field $\psi$.
We show how to derive a result which is free of zero modes, as it should be,
and we explain why the same difficulty does not occur in the case of the square
lattice.
Retaining terms of the same order in the amplitudes $a$, $a^{*}$ as in the 
square lattice case, we find the following effective Hamiltonian
for one hole in the triangular AF (in momentum space)
\begin{eqnarray}
{\cal H}_{\triangle}&=&\frac{3J}{4}\sum_{\bf q}\left\{
(2+w_{\bf q})a^{\dagger}_{\bf q}a^{}_{\bf q}
-\frac{3w_{\bf q}}{2}(a^{\dagger}_{-{\bf q}}a^{\dagger}_{\bf q}
+a^{}_{-{\bf q}}a^{}_{\bf q})\right\}   \nonumber\\
&&-\frac{3JN_s}{8}+\sum_{\bf q}\epsilon_{\bf q} \;h^{\dagger}_{\bf q}h^{}_{\bf q}
    \nonumber \\
&& +\frac{1}{\sqrt{N_s}}\sum_{\bf q,q'}\left\{ f_{\bf q,q'} \;
 h^{\dagger}_{\bf q} h^{}_{\bf q'} \;a_{\bf q-q'} + h.c. \right\} \; .
\label{H3}
\end{eqnarray}
The terms in the first line are nothing but the linear spin wave approximation
(LSW) of the triangular Heisenberg AF. The last line contains the coupling between the hole and the magnons with the hole-magnon vertex
\begin{eqnarray}
f_{\bf q,q'}&=&
  3 \sqrt{3} i \left[ -t\gamma_{\bf q'}+\frac{J}{4}\gamma_{\bf q-q'} \right]
\nonumber \\
&& + \sqrt{3} \left(\frac{1}{2}+w_{\bf q-q'}-i\gamma_{\bf q-q'}\right)
\times\nonumber\\
&&\left[ t (w_{\bf q}-w_{\bf q'}) + \frac{J}{2} (1-w_{\bf q-q'}) \right] \; .
\end{eqnarray}
Here, $w_{\bf q}$ is defined by
\begin{equation}
w_{\bf q}+i\gamma_{\bf q}=
\frac{1}{3} \sum^3_{j=1} e^{i{\bf q} \cdot \mbox{\boldmath $\scriptstyle \delta$}_j}
\end{equation}
and $\mbox{\boldmath $\delta$}_j = (\cos \varphi_j, \sin \varphi_j)$,
$\varphi_j = \frac{2\pi}{3}j$, are the vectors to the three nearest neighbour
sites on the triangular lattice.
As opposed to the square lattice, the hole can hop on the triangular lattice
without emission or absorption of a magnon. This is expressed by the second
line of (\ref{H3}) where
\begin{equation}
\epsilon_{\bf q} = \frac{3J}{4}-3tw_{\bf q} \; ,
\end{equation}
is just half the dispersion of a particle hopping between nearest neighbour
sites of the empty triangular lattice.
${\cal H}_{\triangle}$ constitutes our main result for the Hamiltonian of one
hole in the $t-J$ model on a triangular lattice in spin wave approximation.
In the following, we analyze its content by treating the hole magnon coupling in the simplest approximation.

\subsection{Born approximation of the selfenergy}
\begin{figure}[hbt]
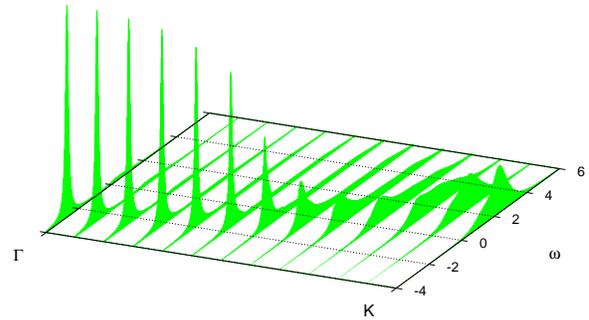

\setlength{\epsfxsize}{\columnwidth}
\centerline{\Bildeins}
\caption[Spectra]{Spectral density $A({\bf k},\omega)$ in SCBA approximation
for $N_s=36 \times 36$ sites, $t=J=1$, and for $\bf k$ values taken along
a straight line from $\Gamma$ to K (cf. Fig. \ref{BZ}) ($\eta=0.1$).}
\label{kspectra}
\end{figure}
Following previous treatments of the square lattice case \cite{Marsiglio91},
we employ the SCBA to obtain the selfenergy $\Sigma({\bf k},\omega)$
of one hole.
We denote the Green's function of one hole by
$G({\bf k},\omega)=(\omega-\epsilon_{\bf k}-\Sigma({\bf k},\omega))^{-1}$.
Then,
\begin{eqnarray}
\Sigma^{SCBA}({\bf k},\omega)\!\!
&=&\frac{1}{N_s}\sum_{\bf k'}
{\left|u_{{\bf k}-{\bf k'}}f_{{\bf k},{\bf k'}} +v_{\bf k-k'}f^{*}_{{\bf k'},{\bf k}}\right|}^2 \times \nonumber\\
&&G^{SCBA}({\bf k'}, \omega- \omega_{\bf k-k'}) \; .
\label{Sigma}
\end{eqnarray}
Here, $\omega_{\bf q}=\frac{3J}{2}\sqrt{(1+2w_{\bf q})(1-w_{\bf q})}$ is the
spin wave energy.
$u_{\bf q}$ and $v_{\bf q}$ are the Bogoliubov amplitudes,
\begin{eqnarray}
u_{\bf q}&=&{\left\{\frac{1}{2}\left[\frac{3J(1+w_{\bf q}/2)}{2\omega_{\bf q}}+1\right]\right\}}^{1/2} \nonumber\\
v_{\bf q}&=&\mbox{sgn}( w_{\bf q}){\left\{\frac{1}{2}\left[\frac{3J(1+w_{\bf q}/2)}{2\omega_{\bf q}}-1\right]\right\}}^{1/2}.\nonumber
\end{eqnarray}
From the Green's function, the spectral density is obtained as
$A({\bf k},\omega)=-\frac{1}{\pi} \lim_{\eta \to 0^+}
\mbox{Im}\,G({\bf k}, \omega+i\eta)$.
The parameter $\eta$ regularizes the spectral density.
\end{multicols}
\widetext
\renewcommand{\thefootnote}{\fnsymbol{footnote}}
\begin{table}[bt]
\begin{tabular}{||c||c|c||c||c|||c|c||c||cc||}
&\multicolumn{4}{c|||}{exact diagonalization}&\multicolumn{4}{c}{linear spin wave, SCBA}& \\ \hline
&\multicolumn{2}{c||}{$t=0$}&$t=0.1$&$t=1$&\multicolumn{2}{c||}{$t=0$}&$t=0.1$&$t=1$&\\ \hline
&$n_h=0$&$n_h=1$&$n_h=1$&$n_h=1$&$n_h=0$&$n_h=1$&$n_h=1$&$n_h=1$&\\
$N_s$&$E^{ex}_{0}$&$E^{ex}_{1,loc}$&$\delta E^{ex}_{1}$&$\delta E^{ex}_1$&$E^{LSW}_{0}$&$E^{SCBA}_{1,loc}$&$\delta E^{SCBA}_1$&$\delta E^{SCBA}_1$&\\ \hline\hline
9 & -5.25000&0.50000&0.40000&4.00000& -5.62500&0.66884&0.31627&3.18974&\\
12& -7.32396&1.15197&0.38544&4.09035& -7.25658&0.64520&0.32854&3.39377&\\
21&-11.78091&0.56862&0.39986&4.06326&-11.82919&0.65145&0.32504&3.35476&\\
24\footnotemark[2]&-12.93870&0.92897&0.29581&3.42937&-13.04174&0.65227&0.32561&3.36573&\\
27&-15.12597&0.59201&0.38805&3.99864&-15.04634&0.65052&0.32572&3.36510&\\
\end{tabular}
\caption{Exact and approximate groundstate energies of samples of $N_s$ sites
of the $t-J$ model with $n_h(=0,1)$ holes. $E_0$: groundstate energies in the half filled case. $E_{1,loc}$: energies for creating a localized hole. 
$\delta E_1$: energy gain of the hole due to delocalization.} 
\label{groundstate}
\end{table}
\begin{multicols}{2}
\narrowtext
\footnotetext[2]{For $N_s=24$ the crystal momentum of the one hole groundstate is
an interior point close to K of the Brillouin zone. This is the reason,
why for $N_s=24$ the values of $\delta E^{ex}_1$ deviate significantly from
the rest.}
\renewcommand{\thefootnote}{\arabic{footnote}}
(\ref{Sigma}) has been solved by numerical iteration starting with $\Sigma =0$ 
for a lattice of $N_s=36 \times 36$ sites with periodic boundary conditions.
In Fig.~\ref{kspectra} we display for $t=1$, $J=1$ the spectral density
$A({\bf k},\omega)$ on $\bf k$--points along a straight line from the centre
$\Gamma$ to the corner K of the Brillouin zone (cf.~Fig.~\ref{BZ}).
The main feature of the spectra is a pronounced quasiparticle peak.
For $t=0$, the quasiparticle dispersion is flat throughout the Brillouin zone.
For increasing $t$, a dispersion with the minimum at $\Gamma$ emerges and the
peak broadens.
Around $\Gamma$, the quasiparticle peak persists for values of $t$ up to $10J$;
near $K$ it already decays for $t>0.2J$.

\subsection{Comparison with exact numerical diagonalization results}
As we have pointed out in the course of the derivation, the effective
Hamiltonian ${\cal H}_{\Delta}$ is an approximate result, and the computation
of the one-hole Green's function is based on the self-consistent Born
approximation. In order to have a check on the validity of these
approximations, we computed the spectral density $A({\bf k},\omega)$ and
related quantities directly
for the $t-J$ model on small periodic samples of the triangular lattice
with $N_s$ sites. Numerically exact results for the quantities in question
were obtained by applying the Lanczos technique (see e.g. \cite{Dagotto94})
to the Hamiltonian of these samples.
In these computations, the exchange constant $J$ has been set equal to unity
throughout. Results for the groundstate energies for half filling ($n_h=0$) and
for one hole ($n_h=1$) are displayed in Table \ref{groundstate} together with 
the approximate LSW and SCBA energies for the same sample sizes.
For $t\neq 0$, the groundstate wave function of the hole is an extended state
with crystal momentum at the high symmetry points of the Brillouin zone;
$\Gamma$ for odd $N_s$ and K for even $N_s$.

\begin{figure}[t]
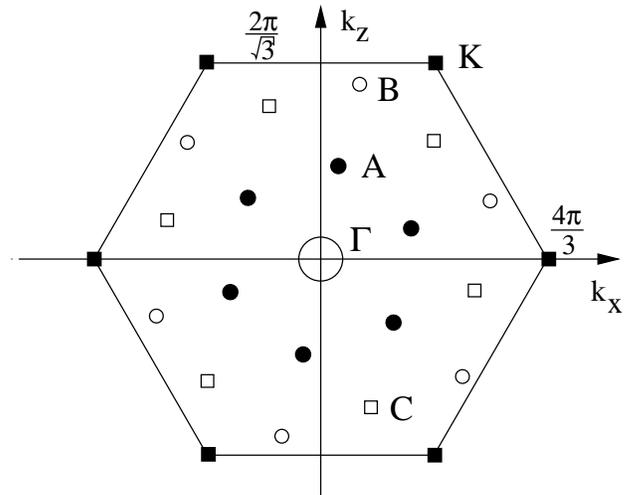

\begin{minipage}[t]{0.95\columnwidth}
\setlength{\epsfxsize}{\columnwidth}
\centerline{\Bildzwei}
\caption[Brillouinzone]{The Brillouin zone of the $N_s=21$ system shows
a sixfold symmetry.}
\label{BZ}
\end{minipage}
\end{figure}
For the half filled case, the $t-J$ Hamiltonian reduces to the Heisenberg
Hamiltonian, and it is well known that in this case the LSW results $E^{LSW}_0$
are in good agreement with the exact results $E^{ex}_0$, in particular for the
larger system sizes \cite{Bernu94}. Turning to the one-hole results, we first
discuss the case $t=0$ in which the hole is localized. $E_{1,loc}$ is the
energy needed to create a localized hole in the half filled groundstate. 
There is a strong even-odd staggering in the exact results (2nd column of Table
\ref{groundstate}): in the even samples ($N_s$ even) in which the spins pair 
up to a singlet, $S_{tot}=0$, the 'binding energy' per electron is larger
than in the odd systems which contain an unsaturated spin, $S_{tot}=1/2$.
This is a finite size effect which, as is seen in Table \ref{groundstate},
decreases very slowly with the sample size. Nevertheless, the exact values 
$E^{ex}_{1,loc}$ appear to converge to the SCBA value $E^{SCBA}_{1,loc}$ in an
alternating fashion. That the even-odd staggering is absent from
$E^{SCBA}_{1,loc}$ is understandable, since the SCBA is based on the spin wave
spectrum of the infinite sample.

\end{multicols}
\widetext
\begin{figure}
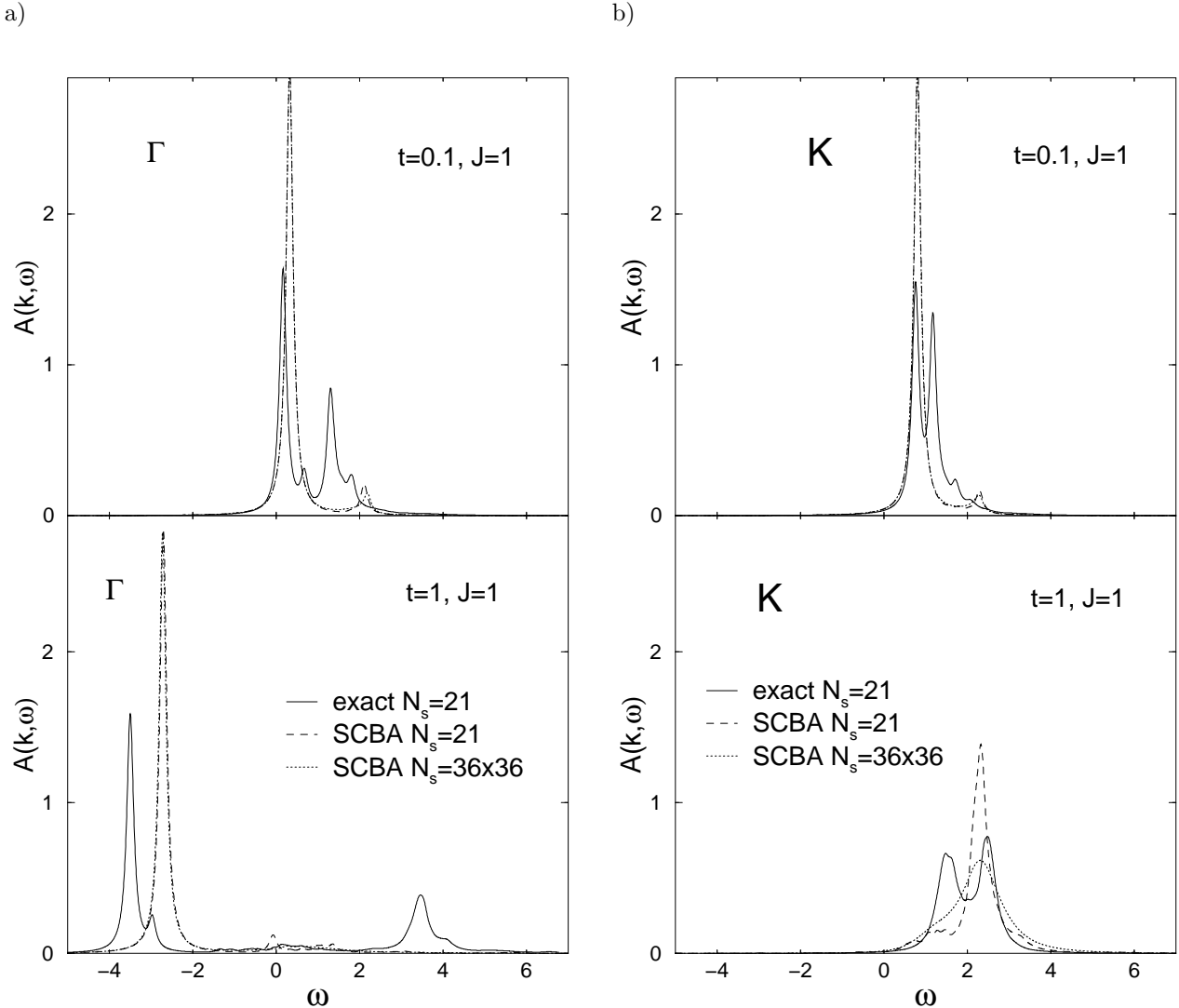

\begin{minipage}[b]{0.95\columnwidth}
\epsfxsize=0.49\columnwidth
\begin{minipage}[t]{\epsfxsize}
a)\\
\centerline{\Bilddreia}
\end{minipage}
\hfill
\epsfxsize=0.49\columnwidth
\begin{minipage}[t]{\epsfxsize}
b)\\
\centerline{\Bilddreib}
\end{minipage}
\caption[Exact and SCBA-Spectra]{Spectral density a) at the $\Gamma$ point
and b) at the corner K of the Brillouin zone for $t=0.1$ and $t=1$ for $J=1$.
The solid line shows the exact results for the $N_s=21$ sites, the dashed
(dotted) line the result of the SCBA for $N_s=21$ ($N_s=36 \times 36$) sites.
$\omega$ is measured against $E^{ex}_{0}$ and $E^{LSW}_{0}$ ($\eta=0.1$).
\label{spectra2}}
\end{minipage}
\end{figure}
\begin{multicols}{2}
\narrowtext
For finite hopping amplitude $t\neq 0$, the
hole becomes delocalized. It is suggestive to write the hole energy as 
$E_1=E_{1,loc}-\delta E_1(t)$, where $\delta E_1(t)$ is the energy gain due to
delocalization of the hole wave function. The results in Table 
\ref{groundstate} show that $\delta E_1$ increases linearly with $t$,
$\delta E_1(t) = \alpha t$. This assertion has been verified by computing
$\delta E_1(t)$ for a sequence of values $0.1 <t<1$ not shown in Table
\ref{groundstate}. The coefficient $\alpha$ is seen to differ significantly
between the exact results and the SCBA, $\alpha^{ex}\approx 4$, 
$\alpha^{SCBA}\approx3.36$. At present, we are not in a position to decide whether
this discrepancy is inherent in the effective Hamiltonian
${\cal H}_{\triangle}$ or whether it is a deficiency of the SCBA.

Next we compare the exact and the SCBA results for the spectral density
$A({\bf k},\omega)$. Fig.~\ref{spectra2} shows the spectral density of one
hole in the $N_s=21$ sample at the centre ($\Gamma$) and at the corner (K) of
the Brillouin zone (cf. Fig.~\ref{BZ}) for the parameter values  $J=1$ and
$t\!=\!0.1$, $t=1$. The solid lines are the results of the exact diagonalization
which were  obtained from a continued fraction expansion \cite{Haydock72} of
the Lanczos results. The dashed lines are the result of the SCBA for $N_s=21$.
For $t\!=\!0.1$ there is a well defined quasiparticle peak in the exact results
at the bottom of the spectrum. This peak is nicely reproduced by the SCBA.

For $t\!=\!1$, a quasiparticle peak is still visible at $\Gamma$ while
one finds a broad structure at the corner of the Brillouin zone. The SCBA
reproduces the peak in the spectrum at $\Gamma$; however, its
position is shifted to higher energies. At K, the SCBA spectrum
shows a structure which becomes broader when the system size is increased from
$N_s\!=\!21$ to $N_s\!=\!36 \times 36$ (dotted lines).
On the other hand, in all cases where we find a quasiparticle peak, the SCBA
results for $N_s\!=\!21$ and for $N_s\!=\!36\times 36$ are indistinguishable.
Thus we conclude that for $t=1$ the quasiparticle peak disappears if one
changes the wavevector from the $\Gamma$ point to the corner of the Brillouin zone, cf. Fig.~\ref{kspectra}.
While there remain differences, we find that overall the SCBA reproduces the
qualitative features of the exact results.
\begin{figure}[tb]
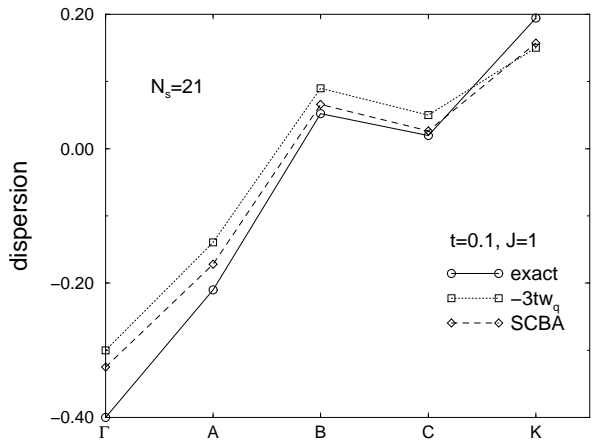

\setlength{\epsfxsize}{\columnwidth}
\centerline{\Bildvier}
\caption[Dispersion]{The dispersion of one hole as defined in the text for $t=0.1$, $J=1$ and $N_s=21$
sites. The symbols $\circ$ show the exact results, the diamonds $\diamond$ the
results of the SCBA and the squares $\Box$ the lowest order approximation for
the Green's function.}
\label{dispersion}
\end{figure}

Finally we discuss the energy dispersion of one hole. We define the dispersion
as the difference of the position of the maximum of $A({\bf k},\omega)$
for a given ${\bf k}$ and the one hole creation energy $E^{SCBA}_{1,loc}$ (see
Table \ref{groundstate}), i.e. the position of the
maximum for $t=0$. In Fig.~\ref{dispersion}, we show the dispersion for 
$N_s\!\!=\!\!21$ sites for $t=0.1$ and $J=1$. The circles (connected by solid
lines) show the results of the exact diagonalization. They follow closely the
free dispersion of the hole, $\epsilon_{\bf q}-3J/4=-3\,t\,w_{\bf q}$
(shown as boxes). The SCBA (diamonds) yields an improvement towards the 
exact results for all points of the Brillouin zone. Obviously, the free
dispersion is already a good approximation of the exact results.

\section{Summary}
In this work, we have developed a formalism which allows the derivation of 
effective Hamiltonians for the motion of holes in an arbitrarily ordered spin
background. Fundamental to this development was the construction of a coherent
state approximation for the electron states, which automatically excludes double
occupancy. In this representation, the partition function of an arbitrary number
of holes can be cast into the form of a path integral which still contains the
full non-linear coupling of the holes to the spin degrees of freedom of the
background. As a first application of this formalism, we have rederived the 
well known effective Hamiltonian ${\cal H}_{\Box}$ for the motion of a single
hole in the collinear antiferromagnetic order of the square lattice
antiferromagnet in the linear spin wave approximation. The derivation of the
effective
\begin{figure}[bt]
\setlength{\epsfxsize}{\columnwidth}
a)\hspace{4.5cm}b)\\
\centerline{\Bildfunf}
\caption{Spin arrangement of $|\Phi_{\bf r}\!>$ (top) and $c^{}_{{\bf r} + \mbox{\boldmath
$\scriptstyle \delta$} \uparrow} c^{\dagger}_{\bf r \uparrow} |\Phi_{\bf r}\!>$ (bottom) for a) the square lattice AF and b) the triangular AF.
\label{config}}
\end{figure}
Hamiltonian
${\cal H}_{\triangle}$ of a single hole moving in the planar spin arrangement
of the triangular antiferromagnet also makes use of the linear spin wave  
approximation. However, in this case, special care has to be taken to ensure a
proper treatment of the Goldstone modes which must not couple to the hole.
In contrast to the square lattice AF, where hopping of the hole from a lattice
site ${\bf r}$ to a neighbouring site {\boldmath $r \unboldmath{+}\delta$} is necessarily 
accompanied by a {\em spin flip} at site $\bf r$, the same hopping process
requires only a {\em rotation of the spin} at ${\bf r}$ by 
$\pm{120}^{\small o}$ in the case of the triangular AF, see Fig \ref{config}.

In other 
words, if $|\Phi^{(0)}_{\bf r}\!>$
is the groundstate with one hole at ${\bf r}$ in
the undistorted spin background of the AFs, then the hopping matrix element
$<\!\Phi^{(0)}_{{\bf r} +\mbox{\boldmath $\scriptstyle \delta$}}\!|c_{{\bf r}+\mbox{\boldmath $\scriptstyle \delta$} \sigma}c^{\dagger}_{{\bf r}\sigma}|\Phi^{(0)}_{\bf r}\!>$ is always zero for the square lattice AF, while it is $+\frac{1}{2}$ for
the triangular AF. This accounts for the main difference between the hole
dynamics in the collinear square lattice AF and the planar triangular AF.

On the square lattice, the hole can hop due to the zero point fluctuations of
the AF background. In this case, magnon-assisted hopping happens
predominantly between the sites of the same magnetic sublattice. This leads to
a minimum of the hole dispersion at a quarter of a reciprocal lattice vector
of the square lattice. By contrast, on the triangular lattice, hole hopping
between nearest neighbour sites, i.e. between different magnetic sublattices,
is also possible without magnon assistance. Thus, for $J>>t$, where magnon
assisted processes are energetically suppressed, the hole dispersion in the
triangular AF will be dominated by the bare (unassisted) hopping processes.
This has been confirmed by comparing the exact dispersion of the bare hole in
the 21 site sample for $t=0.1 J$. The approximate inclusion of the
magnon-assisted processes in the calculation of the dispersion by the
self-consistent Born approximation has been found to improve the agreement
with the exact dispersion.
The exact one--hole Lagrangian (\ref{Lkin}-\ref{Lj}), evaluated in spin wave
approximation, leads to results in reasonable agreement with exact numerical
diagonalization of small systems. ${\cal L}$ provides thus a firm basis for a
derivation of a continuum model which is capable of describing the full 
non--linearities of the spin fields in the presence of single holes.
\end{multicols}
\widetext

\begin{appendix}
\begin{multicols}{2}
\narrowtext
\section{}
A straightforward expansion of $\cal L$ around the classically ordered state
with respect to the fields $\theta$ and $\phi$, and a subsequent integration
of $\psi$ leads to the appearance of zero energy spin wave amplitudes
in linear order in the hopping term ${\cal L}_t$.
This is in contradiction to the spin rotational invariance of the model
(\ref{t-J}).
In order to resolve the contradiction, we now verify that this invariance
persists in an appropriate spin wave expansion.
In the first order expansion around the state
given by $\phi^{}_{\bf r}=\phi^{(0)}_{\bf r}$, $\theta_{\bf r}=\frac{\pi}{2}$
(cf.~\ref{coorsq}), the critical term ${\cal L}_t$ reads
\begin{eqnarray}
{\cal L}_t = && -t \sum_{<\!{\bf r}, {\bf r'}\!>}
 \left\{ \eta^*_{\bf r'} \eta^{}_{\bf r^{}}
 e^{-\frac{i}{2}(\psi_{\bf r'}-\psi_{\bf r^{}})}
 \left[
\cos(\frac{\phi^{(0)}_{\bf r^{}}-\phi^{(0)}_{\bf r'}}{2})
 \right. \right.  \nonumber \\
&& \left. - \sin(\frac{\phi^{(0)}_{\bf r^{}}-\phi^{(0)}_{\bf r'}}{2})
 \left( i \frac{p_{\bf r}+p_{\bf r'}}{\sqrt{2}}
       + \frac{x_{\bf r}-x_{\bf r'}}{\sqrt{2}} \right) \right]\nonumber\\
&& \left. +  \;\;{\bf r} \leftrightarrow  {\bf r'}   \right\}            \; .
\label{Lt1st}
\end{eqnarray}
This expression is valid for both the square, and the triangular lattice.
We now perform an infinitesimal homogeneous rotation in spin space
parametrized by the angles {\boldmath $\epsilon$}.
The corresponding transformation of the fields can be read off from (\ref{coh}).
The field $\eta$ is unchanged and the linear changes in $\psi$, $\theta$, and
$\phi$ are (we express the last two fields by x and p)
\begin{eqnarray}
\psi_{\bf r}' &=&\psi_{\bf r}+2\left[ \epsilon^{x} \cos \phi^{(0)}_{\bf r} +
\epsilon^{y} \sin \phi^{(0)}_{\bf r}\right]\nonumber\\
x_{\bf r}' &=&x_{\bf r} +\sqrt{2}\epsilon^{z}\nonumber\\
p_{\bf r}' &=&p_{\bf r} +\sqrt{2}\left[-\epsilon^{x} \sin\phi^{(0)}_{\bf r}+
\epsilon^{y} \cos\phi^{(0)}_{\bf r}\right].
\label{inftf}
\end{eqnarray}
Substituting these rotated fields for the original ones in (\ref{Lt1st}),
one sees that $\epsilon^{z}$
drops out and the change
in the field $p_{\bf r}$ induced by the other two rotations 
$\epsilon^{x}$ and $\epsilon^{y}$ is compensated
by the corresponding change in the field $\psi_{\bf r}$.
Thus, we see that ${\cal L}_t$, and hence the whole Lagrangian, is invariant
under an infinitesimal rotation in spin space as it should be.
It is only after the integration over the field $\psi$ that the invariance
appears to be lost.

Turning to the case of the square lattice, where 
$\phi^{(0)}_{\bf r} = \pm \pi / 2$ ($\cos\phi^{(0)}_{\bf r} =0$),
the zero modes are seen to decouple in $\psi_{\bf r}$ and $p_{\bf r}$ (cf. (\ref{inftf})), and in ${\cal L}_t$
the term of zeroth order in the fields $x_{\bf r}$ and $p_{\bf r}$ vanishes.
It is easily verified from the spin wave expansion of ${\cal L}_J$
that the zero modes are given by
$x_{\bf r} = x$, $p_{\bf r} = 0$ and
$x_{\bf r} = 0$, $p_{\bf r} = p \sin\phi^{(0)}_{\bf r}$
which both cancel in ${\cal L}_t$.
After substituting $\eta \,e^{\frac{i}{2}(\psi - \phi^{(0)})}$ by $h$,
we can integrate over $\psi$, and return to the operator form.
Thus, the spin wave expansion as sketched in the main text leads to
the correct result \cite{Marsiglio91}.

In the case of the triangular lattice, where
$\phi^{(0)}_{\bf r}=\frac{2\pi}{3},0,-\frac{2\pi}{3}$ for the three
sublattices respectively, the zero modes in $\psi_{\bf r}$ and $p_{\bf r}$ mix.
Again, it is easily verified from the spin wave expansion of ${\cal L}_J$
that they are given by $x_{\bf r} = x$, $p_{\bf r} = 0$, and
$x_{\bf r} = 0$, $p_{\bf r} = p \sin\phi^{(0)}_{\bf r}$, and $x_{\bf r} = 0$, $p_{\bf r} = p \cos\phi^{(0)}_{\bf r}$.
The first mode cancels in ${\cal L}_t$, but the last two remain, if one
disregards $\psi$.
However, we are free to apply a canonical transformation to the fields $\psi$,
$x$, and $p$ in the path integral. We choose
\begin{eqnarray}
\psi_{\bf r}&\to& \psi_{\bf r}+\sqrt{\frac{2}{3}}\left(p_{\bf r}+\frac{2}{3}
\sum_{j=1}^3p_{{\bf r}+{\mbox{\boldmath $\scriptstyle \delta$}_j}}\right)\nonumber\\
x_{\bf r}& \to & x_{\bf r} +\sqrt{\frac{1}{6}}\left(\eta^{*}_{\bf r}\eta^{}_{\bf r} + \frac{2}{3}\sum_{j=1}^{3} \eta^{*}_{{\bf r}-{\mbox{\boldmath $\scriptstyle \delta$}_j}}\eta^{}_{{\bf r}-{\mbox{\boldmath $\scriptstyle \delta$}_j}}\right)\nonumber\\
p_{\bf r}&\to&p_{\bf r}
\label{A3}
\end{eqnarray}
($\mbox{\boldmath $\delta$}_j = (\cos \varphi_j, \sin \varphi_j)$,
$\varphi_j = \frac{2\pi}{3}j$ are the three nearest neighbour vectors.)
With this choice ({\it i}) the new field $\psi$
remains invariant under homogeneous rotations in spin space, i.e. it does not
contain a zero mode, and ({\it ii}) ${\cal L}_{kin}$ is changed only by a
total time derivative which yields no change in the action
(the latter condition is equivalent to the transformation being canonical).
After applying (\ref{A3}) to ${\cal L}$, all zero modes cancel in ${\cal L}_t$.
We substitute $\eta \,e^{\frac{i}{2}(\psi - \phi^{(0)})}$ by $h$,
integrate over $\psi$, and return to the operator representation.
The result for ${\cal H}_{\triangle}$ is the expression (\ref{H3}).

\end{multicols}
\widetext
\end{appendix}

\begin{multicols}{2}
\narrowtext

\end{multicols}
\end{document}